\pdfoutput=1
\documentclass[twocolumn,prb]{revtex4-1}
\usepackage{amssymb,amsmath,bm}
\usepackage{graphicx}%
\usepackage{dcolumn}
\usepackage{color}
\usepackage{longtable}
\usepackage[dvips]{epsfig}
\usepackage[perpage,symbol*]{footmisc}
\usepackage{multirow}
\usepackage{verbatim}
\usepackage{tablefootnote,threeparttable}
\usepackage{braket}
\usepackage{comment}
\usepackage{simpler-wick} 

\usepackage[T1]{fontenc} 

\usepackage{filecontents} 

\usepackage{xr}
\newcommand*{\citen}{}
\DeclareRobustCommand*{\citen}[1]{%
  \begingroup
    \romannumeral-`\x 
    \setcitestyle{numbers}%
    \cite{#1}%
  \endgroup
}

\newcolumntype{.}{D{.}{.}{-1}}
\newcommand{\mc}[3]{\multicolumn{#1}{#2}{#3}}

\newcommand{\fns}{\footnotesize}

\newcommand{\fig}[2]{\scalebox{#1}{\includegraphics{#2}}}

\usepackage{bbm}

\usepackage{hyperref}
\hypersetup{colorlinks,breaklinks,linkcolor=Blue,urlcolor=Blue,anchorcolor=Blue,citecolor=Blue}
\usepackage{color}
\definecolor{DarkBlue}{rgb}{0.0,0.08,0.45}
\definecolor{Blue}{rgb}{0.0,0.0,1.0}
\definecolor{Red}{rgb}{1.0,0.0,0.0}
\definecolor{RedOrange}{rgb}{0.9,0.0,0.2}
\definecolor{dgrn}{RGB}{0,150,0}
\definecolor{dgray}{gray}{0.3}

\usepackage{titlesec}  
\titleformat{\section}[hang]{\bfseries\large}{\thesection}{1em}{}
\renewcommand{\thesection}{\arabic{section}}
\renewcommand{\thesubsection}{.\arabic{subsection}}
\renewcommand{\thesubsubsection}{.\arabic{subsubsection}}
\titleformat{\section}{\normalfont\bfseries}{\thesection.~}{0.25em}{}
\titleformat{\subsection}[runin]{\normalfont\bfseries}{\thesection\thesubsection.}{0.25em}{}
\titleformat{\subsubsection}[runin]{\normalfont\bfseries}{\thesection\thesubsection\thesubsubsection.}{0.25em}{}

\begin{document}
\title{
Diagrammatic Simplification of Linearized Coupled Cluster Theory
}

\author{
    Kevin Carter-Fenk
}
\email[Department of Chemistry, University of Pittsburgh, Pittsburgh, Pennsylvania 15218, USA; ]{kay.carter-fenk@pitt.edu}

\date{\today}


\begin{abstract} 
\begin{center}
    \textbf{Abstract}
\end{center}
\noindent Linearized Coupled Cluster Doubles (LinCCD) often provides near-singular energies in small-gap systems that exhibit static correlation. This has been attributed to the lack of quadratic $\hat{T}_2^2$ terms that typically balance out small energy denominators in the CCD amplitude equations. Herein, I show that exchange contributions to ring and crossed-ring contractions (not small denominators {\em per se}) cause the divergent behavior of LinCC(S)D approaches. Rather than omitting exchange terms, I recommend a regular and size-consistent method that retains only linear ladder diagrams. As LinCCD and configuration interaction doubles (CID) equations are isomorphic, this also implies that simplification (rather than quadratic extensions) of CID amplitude equations can lead to a size-consistent theory. Linearized ladder CCD (LinLCCD) is robust in statically-correlated systems and can be made $\mathcal{O}(n_{\text{occ}}^4n_{\text{vir}}^2)$ with a hole-hole approximation. The results presented here show that LinLCCD and its hole-hole approximation can accurately capture energy differences, even outperforming full CCD and CCSD for non-covalent interactions in small-to-medium sized molecules, setting the stage for further adaptations of these approaches that incorporate more dynamical correlation.
\end{abstract}
\maketitle

\section{Introduction}
Coupled cluster (CC) theory with double
substitutions (CCD) is the simplest form
of CC that captures electron correlation.\cite{SzaOst82}
There are a host of advantages to linearized
CCD methods (LinCCD) over full CCD,\cite{Ciz66, Bar81}
including reductions in memory demands,
ease of
spin-adapting the LinCCD
wave function (albeit there is no rigorous
wave function in linearized CC approximations),\cite{LiPal96}
and simpler physical interpretation of the
equations.
My research group is particularly interested
in the Hermitian formulation that is offered
by linearized CC methods, as this
can be useful in developing excited-state
theories.\cite{BinCar25}
Hermitian approaches are also quite powerful in
the sense that they satisfy the generalized Hellmann-Feynman
theorem, permitting simpler evaluation of forces
(whereas left-eigenvectors are required in non-Hermitian CC
approaches).\cite{CraSch00}

However, linearized CC approaches often
encounter near singularities in small-gap systems,
affecting the performance of LinCCD away from equilibrium.\cite{TauBar09}
Small orbital-energy gaps are often a qualitative
indicator of static correlation,\cite{HolGil11}
where the near-singular behavior of LinCCD can be further understood
as a deficiency resulting from the lack of quadratic
$\hat{T}_2^2$ terms that fold in higher-order correlation
effects necessary to describe static correlation.
In other words, LinCCD lacks the implicit account
for quadruple excitations that is found in the CCD amplitude
equations, making it unable to counteract small energy
denominators. Consequently,
LinCCSD has been combined with Tikhonov regularization\cite{Han98}
as a means of sidestepping divergences.\cite{TauBar09}
Multireference LinCC approaches have also been developed
to avoid divergences in systems that exhibit
static correlation, albeit at
increased cost.\cite{LyaMusLot12,ShaAla15,JeaShaAla17}
Furthermore, LinCC methods have been applied to capture
additional correlation effects atop geminal
reference states that naturally
incorporate static correlation.\cite{ZobSzaSur13,BogAye15,BogTec17,Boz16}

Beyond LinCCD, there are classes of CC approaches
that attempt to correct errors within single-reference
CC that arise due to static correlation.
These ``addition-by-subtraction'' (ABS) CC methods
take the somewhat paradoxical approach of removing
components of the $\hat{T}$ operator that are found to
be particularly ill-behaved in the face of static
correlation.\cite{Bar24}
Perhaps the most well-known ABS-CC approach is
pair CCD,\cite{LimAyeJoh13}
where only pair double substitution clusters are retained,
leading to an approach that can describe single
and double-bond dissociation.\cite{SteHenScu14,HenBulSte14,BrzBogTec19,Bog21}
Alternatively, it is also possible to decouple
the singlet- and triplet-paired amplitudes in CCD
to achieve similarly well-behaved bond dissociation curves.\cite{BulHenScu15,GomHenScu16}
Though, such singlet\slash triplet-pair couplings
occur through a quadratic term in the CCD equations so
the divergence of LinCCD must be attributable to other factors.

Another flavor of ABS-CC approach restricts the CCD
equations to certain classes of diagrams.
For instance, CCD with only ring diagrams
is equivalent to the particle-hole
random-phase approximation (ph-RPA)\cite{ScuHenSor08,JanLiuAng10,CheVooAge17}
and is especially applicable in the case of the high-density
homogeneous electron gas.
Beyond single-reference approaches, diagrammatic re-summations of the ring diagrams have been recently
applied to extend ph-RPA to the
multireference case.\cite{WanFanLi25}

At the other end of the spectrum, CCD restricted to
ladder diagrams (ladder-CCD) is formally equivalent
to particle-particle RPA (pp-RPA)\cite{PenSteVan13,ScuHenBul13} and is
especially suitable in the limit of the low-density electron gas due to
its explicit account of particle-particle correlations.
Diagrammatic analysis of the ring and ladder CC equations
has revealed that ring and ladder diagrams mainly describe
long and short-ranged correlation effects, respectively.\cite{IrmGalHum19}
These naturally imposed length scales have been leveraged
in combinations of ladder- and ring-CCD via
range-separation techniques,
leading to promising methods for describing systems
that do not fall into either extreme.\cite{SheHenScu14}

In this work, I present an ABS linearized CCD approach
that linearizes the ladder CCD amplitude equations.
By removing the terms associated with ring and crossed-ring
diagrams from LinCCD, the resultant linearized ladder CCD (LinLCCD)
approach avoids the near-singularities encountered in these diagrams.
Furthermore, the isomorphism between LinCCD and configuration
interaction with double substitutions (CID) suggests that LinCCD
equations are not size-consistent.\cite{SzaOst82}
A lack of size-consistency implies that, for well-separated molecular fragments $A$ and $B$,
$E_{A\cup B} = E_A+E_B$ is not satisfied by LinCCD\slash CID.
While quadratic corrections have been added to CID to obtain
CCD equations,\cite{PopHeaRag87,CreHe94a,CreHe94b,HeKraCre96,HeHeCre00,Cre13}
revealing the role of $\hat{T}_2^2$ terms in
size-consistent approaches,
I propose removing ring\slash crossed-ring terms from the LinCCD (or
equivalently CID) equations as an alternative route to obtain a size-consistent,
size-extensive, orbital invariant, and naturally regular method.




\section{Theory}
Throughout this work, I will denote occupied
orbitals as $i,j,k,l\dots$,
virtual orbitals as $a,b,c,d,\dots$,
and general unspecified orbitals as $p,q,r,s\dots$.
The abbreviations $n_v$ and $n_o$ will be used
for the number of virtual orbitals and the number
of occupied orbitals, respectively.
Einstein summation notation is used except
in limited
cases where the summation is explicitly written out.

\subsection{Linearized Coupled Cluster Theory}
LinCCD invokes the approximation that the usual
exponential parameterization of the wave function,
\begin{equation}\label{eq:CCwfn}
    |\Psi_{\text{CC}}\rangle = e^{\hat{T}}|\Phi_{\text{HF}}\rangle
\end{equation}
is Taylor-expanded through first order such that,
\begin{equation}\label{eq:LinCCDwfn}
    |\Psi_{\text{LinCC}}\rangle \approx (1+\hat{T})|\Phi_{\text{HF}}\rangle\;~.
\end{equation}
By retaining only strongly-connected diagrams,
the LinCCD energy can be cast in terms of a Hermitian Hamiltonian,
\begin{equation}\label{eq:StrongConnect}
    E = \langle\Phi_{\text{HF}}|[(1+\hat{T}_2^\dagger)\hat{H}(1+\hat{T}_2)]_{\text{SC}}|\Phi_{\text{HF}}\rangle
\end{equation}
where,
\begin{equation}
    \hat{T}_2 = \sum\limits_{\substack{i>j\\a>b}}t_{ij}^{ab}\hat{a}_a^\dagger\hat{a}_i\hat{a}_b^\dagger\hat{a}_j
\end{equation}
is the usual double-substitution operator and the subscript SC indicates that
only the strongly-connected diagrams are retained.
Connected diagrams are defined as
those whose components are all connected via directed lines.
Strongly connected diagrams are a subclass of connected diagrams
in which -- for operators with dual-space components such as
$\hat{T}^\dagger\hat{H}\hat{T}$ in Eq.~\ref{eq:StrongConnect} -- the
removal of one $\hat{T}^\dagger$ or $\hat{T}$ still results
in a connected diagram. On the other hand, if such a removal results in a disconnected
diagram the term ``weakly connected'' is used.\cite{SzaBar92,Tau08}
Restricting Hermitian linearized CC equations to the subclass of strongly connected
diagrams ensures size-extensivity.

The LinCCD doubles amplitude equations
are,
\begin{equation}\label{eq:LinCCD}
\begin{split}
    0 &= v_{ij}^{ab} - \mathcal{P}_{ij}\big(t_{kj}^{ab}f_i^k\big) +\mathcal{P}_{ab}\big(f_c^at_{ij}^{cb}\big)\\
    &\qquad +\frac{1}{2}t_{kl}^{ab}v_{ij}^{kl} +\frac{1}{2}v_{cd}^{ab}t_{ij}^{cd}\\
&\qquad+\mathcal{P}_{ij}\mathcal{P}_{ab}\big(v_{ic}^{ak}t_{kj}^{cb}\big)
\end{split}
\end{equation}
where $v_{pq}^{rs}$ are antisymmetrized 2-electron integrals
$\langle rs||pq\rangle$
and $\mathcal{P}_{pq} = 1-p\leftrightarrow q$ are index permutation operators.
The first 3 terms (first line) of Eq.~\ref{eq:LinCCD}
are often referred to as the {\em driver} terms
(the latter two of which are responsible for the
energy denominator in perturbation theory).
The fourth and fifth terms (middle line of Eq.~\ref{eq:LinCCD})
are associated with {\em ladder} diagrams,
and the final term (last line) emerges from {\em ring} and {\em crossed-ring}
diagrams.
\begin{figure}
    \centering
    \fig{1.0}{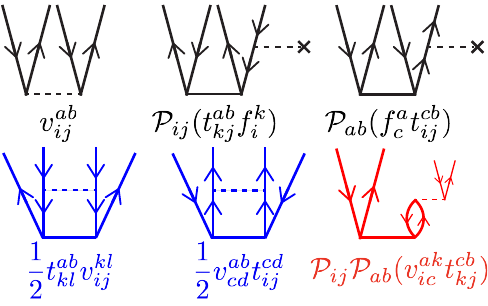}
    \caption{
    Diagrammatic representation of the linearized CCD equations.
    The ``driver'' terms are displayed in black,
    ``ladder'' terms in blue, and ``ring and crossed-ring''
    in red.
    When permutation operations are shown, the associated
    diagram corresponds to the contraction displayed in
    parenthesis.
    }\label{fig:diagrams}
\end{figure}
The LinCCD diagrams are explicitly shown alongside their
corresponding mathematical incarnation in Fig.~\ref{fig:diagrams}.
For more detail pertaining to the CCD (and LinCCD)
diagrams (including the basis set convergence of each term) the reader is referred to Ref.~\citen{MasHumGru24}.
Finally, the LinCCD energy expression in the spatial-orbital basis is explicitly:
\begin{equation}
    E = \frac{1}{4}v_{ab}^{ij}t_{ij}^{ab}
\end{equation}

\subsection{Linearized Ladder Coupled Cluster Doubles}
Interestingly, to my knowledge it has yet to be observed
that retaining only {\em driver} terms and
{\em ladder}-type diagrams within LinCCD leads to
naturally regular equations that are strongly resistant
to divergence in molecular systems.
Notably, linearized ladder approximations
in various combinations with other RPA terms have been explored in
the context of the homogeneous electron gas,\cite{BisLuh78,BisLuh82,Fre83,BisPieSte88a,BisPieSte88b}
but -- to the best of my knowledge -- were essentially abandoned
before being applied in the context of chemistry.
Given the tremendous volume of interest in pp-RPA\slash ladder-CCD
approaches in quantum chemistry\cite{YanPenDav15,YanBurYan15,YanSheZha16,YanDomZha17,AlSSutYan18,BanYuHoh20,YuBanHoh20,LooRom22}
it appears pertinent to explore
linearized ladder approximations.

Applying the same precedent of diagrammatic simplification
as pp-RPA\slash ladder-CCD, I
restrict LinCCD to ladder diagrams yielding,
\begin{equation}
\begin{split}
    0 &= v_{ij}^{ab} + \big(f_c^a\delta_d^b + \delta_c^af_d^b\big)t_{ij}^{cd} - \big(f_i^k\delta_j^l+\delta_i^kf_j^l\big)t_{kl}^{ab}\\
    &\qquad +\frac{1}{2}t_{kl}^{ab}v_{ij}^{kl} +\frac{1}{2}v_{cd}^{ab}t_{ij}^{cd}
\end{split}
\end{equation}
These LinLCCD equations are naturally regular.
In principle, the inclusion of only ladder diagrams
incorporates the most important contributions for
describing strong correlation.
However, to understand precisely how this is the case, it is helpful to notice
that the contractions in terms 2 and 5
and terms 3 and 4 can be grouped together,
\begin{equation}\label{eq:LinLCCD_1}
    \Big(f_c^a\delta_d^b + \delta_c^af_d^b + \frac{1}{2}v_{cd}^{ab}\Big)t_{ij}^{cd} - \Big(f_i^k\delta_j^l+\delta_i^kf_j^l - \frac{1}{2}v_{ij}^{kl}\Big)t_{kl}^{ab} = -v_{ij}^{ab}
\end{equation}

By choosing a clever basis for Eq.~\ref{eq:LinLCCD_1},
such as the one that diagonalizes the
$n_v^2\times n_v^2$ matrix in term 1 and
the $n_o^2\times n_o^2$ matrix in term 2,
the amplitude equation can be reduced to a
highly revealing linear form.
Specifically, one
can define a particle-particle (pp)- and hole-hole (hh)-blocked super Fock matrix with elements,
\begin{subequations}
\begin{equation}\label{eq:SuperFockpp}
    F_{cd}^{ab} = f_c^a\delta_d^b+\delta_c^af_d^b
\end{equation}
\begin{equation}\label{eq:SuperFockhh}
    F_{ij}^{kl} = f_i^k\delta_j^l+\delta_i^kf_j^l
\end{equation}
\end{subequations}
where the eigenvalues of Eq.~\ref{eq:SuperFockpp}, the pp-Fock matrix ($\mathbf{F}_{\rm pp}$),
and Eq.~\ref{eq:SuperFockhh}, the hh-Fock matrix ($\mathbf{F}_{\rm hh}$), are the
hh- and pp-pair energies as estimated by
sums of canonical one-particle orbital energies.
Taking $\mathbf{V}_{\rm pp}$ and $\mathbf{V}_{\text{hh}}$ as matrix representations
of the pp- and hh-integrals from Eq.~\ref{eq:LinLCCD_1},
one can solve the eigenvalue problems,
\begin{subequations}
    \begin{equation}
        \bigg(\mathbf{F_{\rm pp}}+\frac{1}{2}\mathbf{V}_{\rm pp}\bigg)\mathbf{X} = \bm{\lambda}\mathbf{X}
    \end{equation}
    \begin{equation}
        \bigg(\mathbf{F_{\rm hh}}-\frac{1}{2}\mathbf{V}_{\rm hh}\bigg)\mathbf{Y} = \bm{\eta}\mathbf{Y}
    \end{equation}
\end{subequations}
where the eigenvalues
are pp-pair ($\bm{\lambda}$) and hh-pair ($\bm{\eta}$) orbital energies
in the dressed orbital basis, respectively.
They can be decomposed into a one-particle pair contribution
and a two-particle ``dressing'' supplied by the integrals,
\begin{subequations}
    \begin{equation}
        \lambda_{ab} = \tilde{F}_{ab} + \frac{1}{2}\tilde{v}_{ab}^{ab}
    \end{equation}
    \begin{equation}
        \eta_{ij} = \tilde{F}_{ij} - \frac{1}{2}\tilde{v}_{ij}^{ij}
    \end{equation}
\end{subequations}
where $\tilde{F}_{pq}$ are the contributions from one-particle pair energies
and the tilde designates quantities that are in the dressed orbital basis.

Rotation of the supermatrices defined above by eigenvectors
$\mathbf{X}$ in the pp space and $\mathbf{Y}$ in the hh space
leads to a diagonal representation
of Eq.~\ref{eq:LinLCCD_1},
\begin{equation}
    (\lambda_{ab}-\eta_{ij})\tilde{t}_{ij}^{ab} = -\tilde{v}_{ij}^{ab}
\end{equation}
The quantity in parenthesis is diagonal and is thus invertible,
leading to the amplitudes,
\begin{equation}\label{eq:MP2like}
    \tilde{t}_{ij}^{ab} = -\frac{\tilde{v}_{ij}^{ab}}{(\tilde{F}_{ab}+\frac{1}{2}\tilde{v}_{ab}^{ab})-(\tilde{F}_{ij}-\frac{1}{2}\tilde{v}_{ij}^{ij})}
\end{equation}
Here, the elements of $\bm\lambda$ and $\bm\eta$ are expanded to
emphasize that the denominator is constructed of inseparable dressed pair energies.
This distinguishes LinLCCD as a coupled pair theory as opposed to the independent
electron pair theory that constitutes the basis of second-order M{\o}ller-Plesset perturbation theory (MP2),
to which these equations bear a striking resemblance.
Finally, note that this basis transformation was accomplished via independent
occupied-occupied and virtual-virtual orbital rotations, so the
orbital-invariant LinLCCD energy does not change.

Such an isomorphism between MP2
and CCD equations has been noticed before,
but only in the context of mosaic CCD.\cite{SheHenScu14b}
In the linearized ladder context, the hh- and pp-pair energies have been shifted
by the hh and pp integrals,
respectively.
This essentially results in a set of screened
first-order amplitudes wherein the energy gap is
widened by adding hh correlations
to the one-particle occupied pair energies
and pp correlations to
one-particle virtual pair energies, making LinLCCD robust against
divergence in small-gap systems.
Unlike mosaic CCD and myriad other renormalized MP2 theories,\cite{KelTsaReu22,ZhaRinSch16,ZhaRinPer16,CarSheHea23,DitHea25}
the LinLCCD gap is widened in an amplitude-independent way,
suggesting that the LinLCCD equations could be solved non-iteratively,
albeit such an approach would be quite impractical as it would require
the diagonalization of a $n_v^2\times n_v^2$ matrix.

Incredibly, removing the ring and crossed-ring contractions from LinCCD
also corrects the size-consistency errors in the parent method.
I note that beyond size-consistency, LinLCCD is also size-extensive
and orbital invariant.
As the LinLCCD equations can be recast in the form of
Eq.~\ref{eq:MP2like}, the proof for size-consistency
in this basis is trivial.
Consider a system wherein molecules $A$ and $B$ are
sufficiently far apart that their respective molecular
orbitals may be trivially fragment-ascribed.
After a rotation into the aforementioned dressed orbital
basis Eq.~\ref{eq:MP2like} becomes zero for all
sums over disjoint orbitals by means of the electron repulsion integrals in the numerator,
ensuring that the resultant energy satisfies $E_{A\cup B} = E_A + E_B$.



My group is particularly interested in low-scaling approximations
that describe static correlation qualitatively.
While I will demonstrate the utility of LinLCCD, it does
retain the most expensive $\mathcal{O}(n_o^2n_v^4)$ particle-particle
ladder term that is responsible for the $\mathcal{O}(N^6)$
cost of CCD.
In the spirit of exploring low-scaling variants of linear ladder theories,
I introduce one further approximation by completely removing
the costly particle-particle ladder term to achieve,
\begin{equation}\label{eq:LinLCCD_hh}
    \Big(f_c^a\delta_d^b + \delta_c^af_d^b\Big)t_{ij}^{cd} - \Big(f_i^k\delta_j^l+\delta_i^kf_j^l - \frac{1}{2}v_{ij}^{kl}\Big)t_{kl}^{ab} = -v_{ij}^{ab}
\end{equation}
thus shifting only the occupied-pair energies
by the hole-hole ladder term.
Unlike the LinLCCD case, there is no diagrammatic
justification for removing the particle-particle
ladder term, but the resultant
LinLCCD(hh) approach is a potentially fruitful approximation
that scales much more favorably as $\mathcal{O}(n_o^4n_v^2)$.

It should be noted that the LinLCCD(hh) approximation
does not sacrifice orbital invariance, size-consistency, or size-extensivity.
Such claims can be verified by applying analagous
occupied-occupied orbital rotations
to Eq.~\ref{eq:LinLCCD_hh} such that term 2 becomes diagonal.
Once again, this gives way to a set of dressed amplitude equations
that can be written as,
\begin{equation}\label{eq:hh_mp2_amps}
    (\varepsilon_a+\varepsilon_b-\eta_{ij})\tilde{t}_{ij}^{ab} = -\tilde{v}_{ij}^{ab}
\end{equation}
where now only the occupied pair orbital energies have been dressed with hh correlation.
While these basis transformations are revealing as to the underlying nature of
various linearized CC approximations, in practice my implementation
solves Eq.~\ref{eq:LinLCCD_1} self-consistently.

\section{Results \& Discussion}
\subsection{Bond Breaking}

\begin{figure}
    \centering
    \fig{1.0}{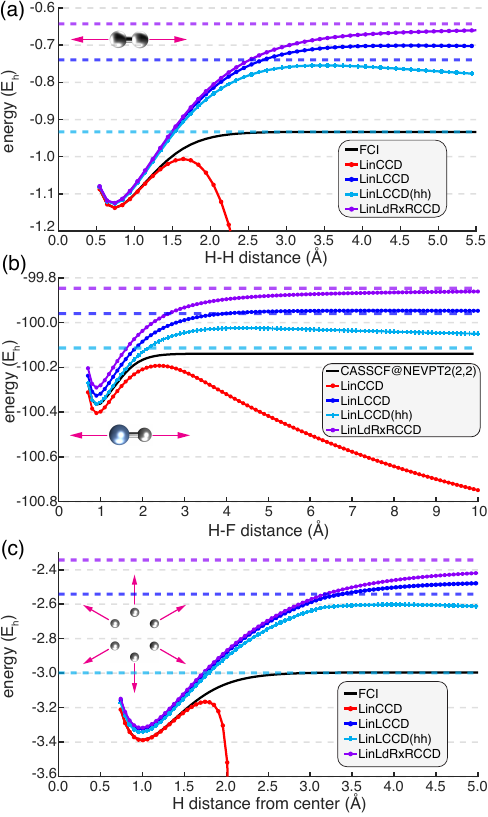}
    \caption{
    Dissociation curves for (a) hydrogen molecule
    in the minimal STO-3G basis set,\cite{HehStePop69} (b)
    Hydrogen fluoride molecule in the aug-cc-pVQZ basis,\cite{Dun89,WooDun94}
    and (c) H$_6$ in the cc-pVDZ basis.
    Dashed lines of like-color indicate the
    dissociation limit for a particular method.
    Dissociation limits were
    estimated at $R=10^6$~\AA, except in the case of
    H$_6$ where $R=10^3$~\AA\ was used instead
    due to convergence difficulties.
    }\label{fig:All_PES}
\end{figure}

I first show the relative robustness of LinLCCD methods through a few simple bond dissociation potential energy surfaces. The simplest case of H$_2$ in the STO-3G basis is shown in Fig.~\ref{fig:All_PES}a.
As expected, LinCCD diverges rapidly as the H--H bond is stretched.
However, both LinLCCD and the further pruned LinLCCD(hh) approaches
smoothly dissociate H$_2$ towards some limit, which is exact in the case of LinLCCD(hh).
Of course, I note that these results also show that
LinLCCD is no longer exact for all two electron systems, but neither is its
(chemically very useful) parent ladder approximation, pp-RPA.

Interestingly, if the linear direct ring and crossed-ring terms ({\em i.e.} without
antisymmetrizing the two-electron repulsion integrals)
are retained alongside the fully antisymmetric ladder terms,
the resultant LinLdRxRCCD approach also does not diverge.
LinLdRxRCCD differs from full LinCCD only by the exchange component
of the linear ring and crossed-ring contractions, implying that the
term most responsible for the instability of LinCCD is not
the small orbital-energy denominator {\em per se}, but the
exchange ring\slash crossed-ring terms.
In the case of HF molecule dissociation in Fig.~\ref{fig:All_PES}b,
this finding helps to explain why regularization by means of eliminating
near-zero denominator terms in LinCCSD is only somewhat effective,\cite{TauBar09}
eventually breaking down
at large $R_{\text{H-F}}$, whereas LinLCCD, LinLCCD(hh),
and LinLdRxRCCD are all stable out to $10^6$~\AA.

I note that others have put forth that
the divergence of CC equations alongside other
deficiencies in statically-correlated systems manifest
due to various exchange terms.\cite{KatMan13,ScuHenSor08,ScuHenBul13,RisPerBar16}
Similar instabilities were also recently reported
for renormalized propagator methods
that include ring and crossed-ring exchange terms.\cite{OpoPawOrt25}
Of course, removing exchange terms can lead to undesirable self-interaction artifacts,
as is well known in the case of direct ring CCD
(otherwise known as the direct ph-RPA),
which over-binds significantly at equilibrium geometries.\cite{MorCohYan12,PaiRenRin12,RuzZhaSch15,RuaRenGou21}
Though less common, removing exchange terms from ladder CCD
has also been explored but appears to lead to less satisfactory results for
bond dissociation energies.\cite{TahRen19}
Thus, if a self-interaction-free theory is desired
that can smoothly dissociate bonds to a clear asymptotic limit, LinLCCD
and LinLCCD(hh) represent suitable options.

Lastly in the series of bond dissociation curves, I investigate hexagonal H$_6$
dissociation in Fig.~\ref{fig:All_PES}c. The hexagonal H$_6$ system is prototypical
of strongly-correlated systems in chemistry and is reminiscent of the Hubbard model
Hamiltonian.
As the H$_6$ ring is expanded LinCCD rapidly diverges while all methods that
exclude ring\slash crossed-ring exchange diagrams remain stable.
The estimated asymptotic limit of LinLCCD(hh) appears somewhat deceptive as plotted because
it very slightly overestimates the correlation energy at large $R$, dipping
below the full configuration interaction (FCI) reference curve by
about 4~mE$_{\text{h}}$. While the LinLCCD(hh) potential curve is far too repulsive at intermediate $R$,
the asymptotic limit remains very impressive for such an approximate scheme.
Overall, all methods that remove the exchange ring\slash crossed-ring terms are robust for
the bond stretching coordinates investigated in Fig.~\ref{fig:All_PES}
while LinCCD fails in all cases, suggesting that ring\slash crossed-ring exchange
contractions are to blame for the near-singular behavior of LinCCD.


\subsection{Non-Bonded Interactions}
Recall that LinLCCD is an approximation to pp-RPA,
so it should not be expected to be a quantitatively
accurate approach for total energies as pp-RPA -- and therefore LinLCCD -- does not offer ideal coverage of dynamical correlation
effects.\cite{PenSteVan13,SheHenScu14b,YanPenDav15}
These deficiencies can be accounted for in pp-RPA by combination with density functionals,
but I will not explore this here.\cite{VanYanYan13,PenVanYan14,YanPenLu14,YanSheZha16,JinYanZha17,AlSSutYan18}
The lack of dynamical correlation can be immediately seen in the potential
energy curves of Fig.~\ref{fig:All_PES}, as
LinLCCD and LinLCCD(hh) both underestimate the magnitude of the correlation energy relative to FCI
and LinCCD near equilibrium.
However, much of the utility in RPA methods comes
from how accurately they predict energy differences
rather than the total energies themselves.\cite{PenSteVan13}
In this section I will explore the accuracy of
LinLCCD for noncovalent interactions computed via,
\begin{equation}
    \Delta E_{\text{int}} = E_{AB} - E_A - E_B\;~,
\end{equation}
where $E_{AB}$ is the energy of the complex and $E_X$ is the monomer energy of fragment $X$.
All interaction energy calculations have been counterpoise corrected.\cite{BoyBer70}

\begin{figure}
    \centering
    \fig{1.0}{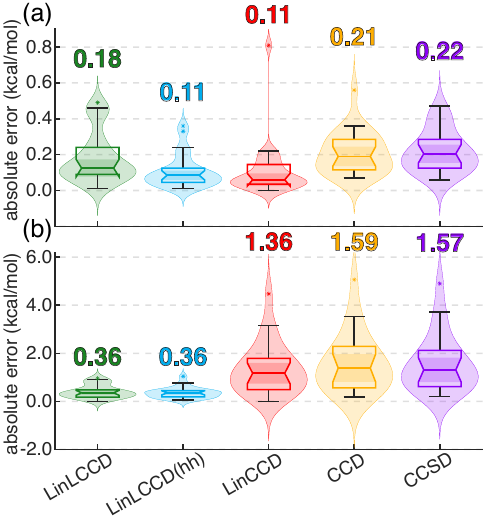}
    \caption{
    Error statistics for noncovalent interaction energies
    extrapolated to the complete basis set limit for
    (a) the A24 set of small dimers and (b)
    the S22 data set of small to medium sized dimers.
    The inset numbers indicate the mean absolute error
    in kcal\slash mol.
    }\label{fig:NCI}
\end{figure}

First, I examine the performance of LinLCCD, LinLCCD(hh),
LinCCD, CCD, and CCSD on the A24 data set of small dimers in
Fig.~\ref{fig:NCI}a.\cite{A24}
These results are largely unremarkable as all methods
perform statistically about the same with relatively low errors
ranging between 0.1--0.2~kcal\slash mol.
Some notable differences are seen in Fig.~\ref{fig:NCI}b for the S22 data set
which features several medium-sized $\pi$-stacked systems.\cite{S22,S22A}
Interestingly, while LinLCCD and LinLCCD(hh) perform similarly with quite low
mean absolute errors (MAE) of 0.4~kcal\slash mol, LinCCD performs poorly
with a MAE of 1.4~kcal\slash mol.
This poorer performance is
in line with the full CCD and CCSD methods, which
also under-perform on S22.
While it is sensible that LinCCD, CCD, and CCSD perform
similarly at equilibrium geometries,
it is reasonable to wonder whether any particular interaction type
(H-bonding, dispersion-bound, $\pi$-stacking, or mixed
interactions) is responsible for the uniformly poor performance
of these methods. However, Figure~\ref{fig:SI_S22} suggests that
the performance of all approaches remains consistent
across interaction types. This is especially
interesting in the case of LinLCCD methods, which
are akin to renormalized MP2. Whereas MP2 tends
to have a significant propensity to over-bind
$\pi$-stacked
complexes by upwards of 100\%,\cite{CarLaoLiu19,NguCheAge20}
no such bias is noted in LinLCCD approaches for the dimers in S22.

Both LinLCCD and LinLCCD(hh) methods appear to perform at least
as well as regularized perturbation theory approaches.\cite{SheLoiHai21,CarHea23}
These results suggest that, despite a lack of dynamical correlation,
the energy differences obtained with LinLCCD approaches
are quite accurate for non-bonded interactions,
surpassing those of full CCD and CCSD.

\begin{figure}[h!!]
    \centering
    \fig{1.0}{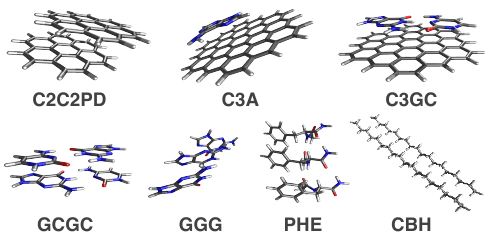}
    \caption{Systems that comprise the L7 data set
    along with their commonly employed acronyms.}
    \label{fig:L7Systems}
\end{figure}

\begin{table}[h!!]
\caption{Interaction Energies for L7 Data Set (kcal/mol)}
\setlength{\tabcolsep}{1pt}
\label{tab:L7}
\begin{tabular}{l ...c}
\hline\hline
\mc{1}{c}{System}   & \mc{1}{c}{LinLCCD(hh)$^a$} & \mc{1}{c}{BW-s2}$^{a,b}$ & \mc{1}{c}{MP2}$^{a,b}$ & \mc{1}{c}{CCSD(T$_0$)$^{a,c}$}\\ 
   \hline
C2C2PD & -34.54 & -33.32  & -38.08 & $-20.93\pm 0.44$ \\
C3A & -25.06 & -24.11 & -27.09 &  $-17.49\pm 0.31$ \\
C3GC & -41.54 & -40.40 & -45.37  &  $-29.24\pm 0.91$\\
GCGC & -16.50 & -17.00 & -18.99  &  $-13.54\pm 0.27$\\
GGG & -3.31 & -3.63 & -4.54 &  \ \ $-2.08\pm 0.09$\\
CBH & -9.40 & -10.90 & -11.83  & $-11.00\pm 0.17$\\
PHE & -25.94 & -25.73 & -26.32 &  $-25.46\pm 0.01$\\
\hline
MAE & 5.36 & 4.78 & 7.18   &  \text{---}\\ 
\hline\hline
\mc{5}{l}{\fns $^a$Extrapolated to CBS limit.}\\
\mc{5}{l}{\fns $^b$From Ref.~\citen{CarSheHea23}.}\\
\mc{5}{l}{\fns $^c$DLPNO-CCSD(T$_0$) from Ref.~\citen{VilBalWan22}.}\\
\end{tabular}
\end{table}

In an effort to emphasize the relative
affordability of the $\mathcal{O}(n_o^4n_v^2)$
LinLCCD(hh) approximation, I also present complete basis set
limit extrapolated interaction energies for the L7
data set of large dimers in Fig~\ref{fig:L7Systems}.\cite{L7}
I compare to the complete basis set limit domain-localized pair natural orbital (DLPNO)\cite{RipSanHan13,RipPinBec16,GuoBecNee18}
CCSD(T$_0$) benchmark data of Lao and co-workers.\cite{VilBalWan22}
I note that DLPNO-CCSD(T$_0$) results
are quite sensitive to the particular thresholds
chosen, so the benchmark data have
some unknown error in addition to those imposed
by the $(T_0)$ correction. This error is roughly 2--6 kcal\slash mol for the $\pi$-stacked
complexes in L7.\cite{GraHer24}

The results in Table~\ref{tab:L7} show that
MP2 dramatically overestimates the interaction energies in L7,\cite{NguCheAge20}
which features several large $\pi$-stacked systems.
As LinLCCD(hh) can be viewed as a renormalized MP2, it is of
interest to contrast its performance with MP2 and with
the renormalized MP2 approach known as
size-consistent Brillouin-Wigner perturbation theory (BW-s2).\cite{CarHea23}
The results in Table~\ref{tab:L7} suggest that including
hole-hole relaxation in the one-particle energies can temper
the overestimated interaction energies of conventional MP2,
but not dramatically so. The MAE is reduced relative to MP2
by nearly 2~kcal\slash mol, which is an improvement but suggests
that linear ladder correlation is insufficient to achieve quantitative
accuracy. The renormalization supplied by BW-s2 is somewhat more effective at suppressing
over-correlation in the largest
$\pi$-stacked systems than LinLCCD(hh), but less aggressive in systems like
GCGC, GGG, and CBH, leading to an overall MAE that is about 0.6~kcal\slash mol
lower than LinLCCD(hh). That said, the results between BW-s2 and LinLCCD(hh)
are comparable to within 12\% of one another.
While I have shown that LinLCCD(hh) is affordable enough to
be applied to such large systems and that the results are somewhat improved relative to MP2,
some empiricism,\cite{CarSheHea23} or other means of incorporating
additional many-body screening effects could beget improvements.

\subsection{Quest \#8 Singlet/Triplet Gaps}
Spin state energetics of transition metal complexes
are a key quantity in the predictive modeling of
energy relevant systems such as
photoredox catalysts that are used to generate solar fuels
such as H$_{\text{2}}$ and to facilitate organic syntheses.\cite{PriRanMac13,AriMcC16,ForHei19,ChaGhoYar23}
While there remains a paucity of benchmark-quality spin-state energetics data
for transition metal complexes in the literature, the small Quest~\#8 data set of Loos
and co-workers does provide excellent data for comparisons with non-relativistic quantum chemistry methods.\cite{QUESTDB,Quest8}
The Quest~\#8 set contains 11 diatomic, monometallic transition metal molecules
with non-relativistic theoretical best estimate (TBE) values computed in the gas phase,
making comparison with other non-relativistic quantum chemistry methods more straightforward
than (still useful) back-corrected experimental values.\cite{SSE17}
Herein, I evaluate the performance of various correlated wave function
theoretic approaches on the singlet\slash triplet gaps of the 7 molecule
subset in Quest~\#8 that has a singlet ground state.

\begin{figure}
    \centering
    \fig{1.0}{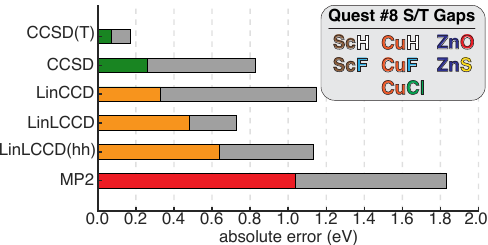}
    \caption{
    Mean absolute errors (shown as colored bars)
    and maximum errors (shown as gray bars) for the lowest energy
    singlet\slash triplet gaps in Quest~\#8 as computed by
    various $\Delta$CC and $\Delta$MP2 methods.
    The bars are colored according to the category of approximation being
    applied, where M{\o}ller-Plesset perturbation theory is red,
    linearized CC methods are orange, and nonlinear CC approaches are green.
    The systems in question are shown in the inset.
    }\label{fig:STGaps}
\end{figure}

MAEs and maximum errors for the  Quest~\#8
singlet\slash triplet gaps are shown in Fig.~\ref{fig:STGaps}.
None of the methods that truncate at double substitutions
are quantitatively accurate relative to the TBE benchmarks,
but the inclusion of perturbative triple excitations in CCSD(T)
clearly has a large effect on the accuracy of the predicted gaps,
reducing the MAE from 0.3~eV with CCSD to
0.07~eV with the inclusion of triples.
As one might expect, LinCCD performs similarly
well, albeit with a larger maximum error of 1.2~eV.
Interestingly, while LinLCCD features a slightly larger MAE
than LinCCD at 0.5~eV, it  has a smaller maximum error
than both LinCCD and CCSD at just 0.7~eV.
Of all of the linearized CC approaches, LinLCCD(hh) performs the worst
with a MAE of 0.6~eV, but it still outperforms MP2.
As the linearized ladder CCD approaches can be conceptualized as intermediate
theories between MP2 and LinCCD, it makes sense that their MAEs fall
in a hierarchical order MP2$>$LinLCCD(hh)$>$LinLCCD$>$LinCCD.
While the results presented here are in line with expectations and are reasonably accurate,
they once again point towards a need for incorporating more dynamical
correlation into the LinLCCD approximation.
Such studies are currently underway in my group.

\subsection{Photolysis of Volatile Organic Compounds}
The previous results focus mainly on energy differences
between the electronic ground state and a high-spin triplet state of
the system.
This is one example of a $\Delta$CC calculation, where the energy of each
electronic configuration is optimized at the self-consistent field level
and used as the reference for a subsequent non-Aufbau CC calculation.
Excitation energies are then obtained by taking the difference between
the CC energies of each respective calculation.
In this section, I will explore the calculation of excited state
bond dissociation curves via the $\Delta$CC approach
to assess LinLCCD for its recovery of potential surfaces of
open-shell systems.

\begin{figure}
    \centering
    \fig{1.0}{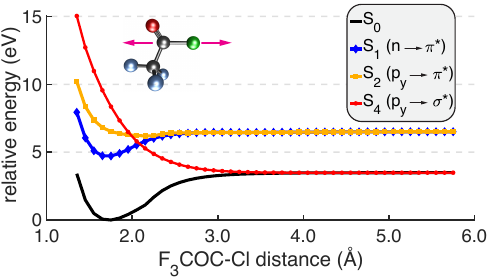}
    \caption{Dissociation of CF$_3$COCl along the C--Cl bond
    using $\Delta$LinLCCD from reference configurations that represent
    the ground electronic state as well as
    $n\rightarrow\pi^\ast$, $p\rightarrow\pi^\ast$, and $p\rightarrow\sigma^\ast$ transitions.}
    \label{fig:VOC}
\end{figure}

Volatile organic compounds are of intense interest in the atmospheric
chemistry community and understanding their photochemistry
can inform on human health\cite{YanZhuLuo23,ZhoZhoWan23} and global climate modeling.\cite{MelWalChe15,LiZhaYin24,HuaZheLi25}
One downstream product of CHCl$_2$F, a recently phased-out refrigerant,
is CF$_3$COCl, which is known to decompose under UV irradiation.\cite{JanVinRak23}
The results in Fig.~\ref{fig:VOC} show $\Delta$LinLCCD calculations
on several of the lowest-energy excited states in CF$_3$COCl
along the C--Cl bond stretching coordinate.
Typically I would include single substitutions within the CC {\em ansatz}
to model open-shell systems, but the presence of singles clusters
threatens the stability of the excited-state configuration.
By Thouless' theorem,\cite{Tho60} single substitutions are equivalent to orbital rotations
that could push the desired excited-state solutions towards the ground state. 
Therefore, a caveat to bear in mind in the following analysis is that the
reference determinant for each $\Delta$LinLCCD excited state is spin contaminated
and without single substitutions to aid in spin purification
the calculated excitation energies are likely underestimated.\cite{PurSekBar88,Sta94,Kry00}
While approximate spin projection could be employed,\cite{Sch88}
I do not expect this to impact the qualitative validity of the results -- especially
for the purposes of modeling the topography of each potential surface
at large C--Cl distances.

The dissociation curves for various states of CF$_3$COCl are shown
in Fig.~\ref{fig:VOC}.
All of the LinLCCD results are qualitatively consistent with
the multireference calculations of Ref.~\citen{JanVinRak23}.
Namely, the $S_1$ state corresponds to a bound $n\rightarrow\pi^\ast$ transition
at 4.7~eV, which is in good agreement with the experimental band maximum
of about 4.9~eV.\cite{JanVinRak23}
Furthermore, the $S_2$ and $S_4$ states are unbound and
correspond to $p\rightarrow\pi^\ast$ and $p\rightarrow\sigma^\ast$ transitions, respectively.
Both states lead to free dissociation to two different limits.
This is expected, because dissociation along the ground state potential surface
should lead to the homolytic cleavage of the C--Cl $\sigma$ bond, populating
the $\sigma^\ast$ orbital and resulting in the same dissociation limit as the
$S_4$ state.
In the case of the $S_2$ state, the occupied $\pi^\ast$ orbital
corresponds to a qualitatively different configuration at dissociation.

The dissociation limits for the ground state and $S_2$ states are 3.5~eV
and 6.5~eV, respectively. The former is very close to the extended multistate complete
active space second-order perturbation theory (XMS-CASPT2)\cite{ShiGyoCel11}
dissociation energy for $S_0$ predicted in Ref.~\citen{JanVinRak23} of 3.44~eV.
The energy difference between $S_0$ and $S_2$ states
at dissociation is 3~eV and is also in excellent agreement
with XMS-CASPT2 results.
Despite spin contamination, the qualitative curvature of each surface is reasonable
and LinLCCD with open-shell singlet references provides excellent
bond dissociation energies in the case of CF$_3$COCl.

\section{Conclusions}
In summary, I have introduced linearized ladder CCD equations
alongside a linear hole-hole ladder approximation and examined
their applicability to a range of chemistry contexts.
LinLCCD and LinLCCD(hh) are both robust when static correlation becomes important,
and LinLCCD(hh) serves as an affordable $\mathcal{O}(n_o^4n_v^2)$ approximation that can be applied to
large systems.
I have also shown that the most problematic terms in LinCCD that lead to near-singular correlation
energies are the ring
and crossed-ring contractions -- particularly the exchange contributions therein.
It is especially notable that LinLCCD is a size-consistent CID method that is obtained
by removing linear terms rather than adding quadratic ones.\cite{PopHeaRag87}
With future adaptations of LinLCCD and LinLCCD(hh) to incorporate more
dynamical correlation, these approaches could prove to be quite useful in the design
of new computational methodologies for modeling strongly correlated systems
reasonably well within single-reference approximations.

\section*{Computational Details}
All calculations make use of the resolution-of-the-identity
approximation and were performed using a developer version of
Q-Chem~v6.2.\cite{QCHEM5}
The A24 and S22 complete basis set limit results were
obtained using two-point aug-cc-pVDZ\slash aug-cc-pVTZ extrapolation
using $\beta=2.51$ and $\alpha=4.3$ as per Neese and Valeev.\cite{NeeVal11}
The S22 calculations made use of frozen natural orbitals (FNOs),
retaining 99.6\% of the natural orbital occupation.\cite{SosGeeTru89,TauBar05,TauBar08}
The use of FNOs has been shown to be a robust approximation that provides
benchmark-quality non-covalent interaction
energies on the S22 set.\cite{DePShe13}
The L7 calculations were extrapolated to the complete basis set limit
using the somewhat smaller Def2-ma-SVP and Def2-ma-TZVP basis sets for heavy
atoms and the corresponding Def2-SVP\slash Def2-TZVP for H atoms.\cite{def2,def2aug,GraHer22}
As Neese and Valeev do not provide parameters to minimize extrapolation errors for
Karlsruhe basis sets with diffuse functions,\cite{NeeVal11} I computed the Hartree-Fock energies for L7
with the Def2-ma-QZVPP basis set and extrapolated the
correlation energy with the more typical $\beta=3$ parameter.
The L7 calculations are the only ones that use the frozen core approximation.
The Quest\#8 calculations employ the aug-cc-pVTZ basis for even-handed comparison with the
aug-cc-pVTZ reference data
in Ref.~\citen{Quest8} (due to limitations in Q-Chem, all
$I$ angular-momentum functions
were removed from the auxiliary basis).
Finally, the CF$_3$COCl potential energy surfaces were calculated using unrestricted
Hartree-Fock reference orbitals. Non-Aufbau reference configurations were stabilized using
a combination of state-targeted energy projection and initial maximum overlap method algorithms.\cite{CarHer20b,GilBesGil08,BarGilGil18b}

\section*{Acknowledgements}
I thank Sylvia J. Bintrim and Abdulrahman Y. Zamani for our many engaging
and enlightening discussions.
This research was supported in part by the University of Pittsburgh
and the University of Pittsburgh Center for Research Computing, RRID:SCR\_022735, through the resources provided.
Specifically, this work used the H2P cluster, which is supported by NSF award number OAC-2117681.

\section*{Supporting Information}

The Supporting Information is available free of charge at https://pubs.acs.org/doi/10.1021/acs.jpca.XXXXXXX
\begin{itemize}
\item[] All A24 and S22 data, finite-basis set L7 results,
and Quest\#8 data (XLSX)
\item[] Additional analysis of interaction subtypes in
the S22 data set (PDF)
\end{itemize}


\providecommand{\refin}[1]{\\ \textbf{Referenced in:} #1}


\clearpage
For table of contents only:\\
\begin{center}
TOC Graphic\\
\fig{1.0}{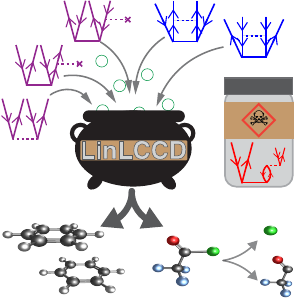}
\end{center}

\end{document}


\title{
Supporting Information for 
\\
``Diagrammatic Simplification of Linearized Coupled Cluster Theory''
}

\author{
    Kevin Carter-Fenk*\footnote[0]{*kay.carter-fenk@pitt.edu}
}
\affiliation{
	Department of Chemistry, University of Pittsburgh, Pittsburgh, Pennsylvania 15218, USA
}

\date{\today}

\maketitle
\thispagestyle{plain}

\section*{S22 Decomposition of Interaction Types}
\begin{figure}[h!!]
    \centering
    \fig{1.0}{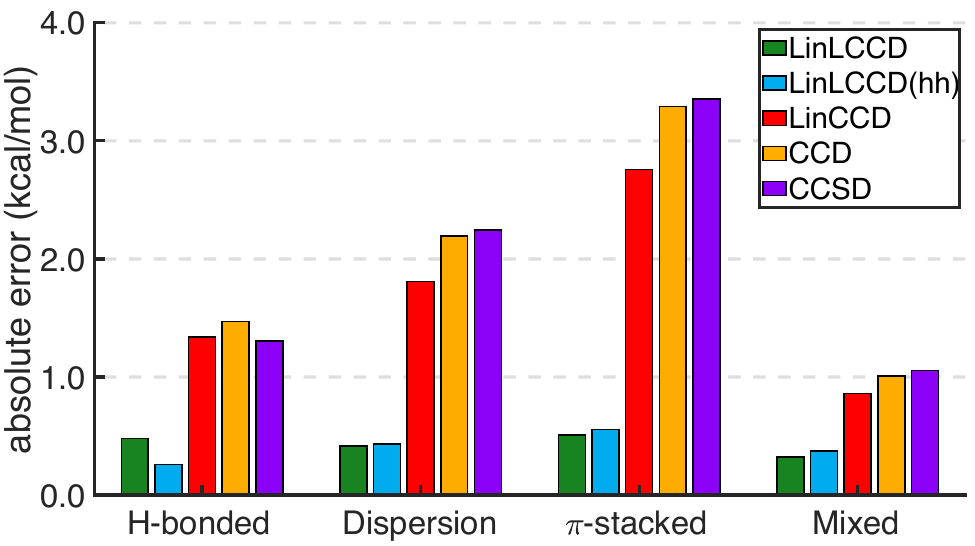}
    \caption{Mean absolute errors across
    various non-covalent interaction types in
    the S22 data set.}\label{fig:SI_S22}  
\end{figure}